\documentclass[12pt]{article}
\title{\textbf{PTAS for $k$-tour cover problem on the plane for moderately large values of~$k$}
    \thanks{Research supported in part by the Centre for Discrete
        Mathematics and its Applications (DIMAP), EPSRC award EP/D063191/1,
        and by VR grant 621-2005-408.}\\
}

\author{\textbf{Anna Adamaszek}
    \thanks{Department of Computer Science and
    Centre for Discrete Mathematics and its Applications (DIMAP),
    University of Warwick.
    Email: A.M.Adamaszek@warwick.ac.uk.}
        \and
    \textbf{Artur Czumaj}
    \thanks{Department of Computer Science and
    Centre for Discrete Mathematics and its Applications (DIMAP),
    University of Warwick.
    Email: A.Czumaj@warwick.ac.uk.}
        \and
    \textbf{Andrzej Lingas}
    \thanks{Department of Computer Science, Lund University.
    Email: Andrzej.Lingas@cs.lth.se.}
}

\author{\textbf{Anna Adamaszek} \\
    \normalsize DIMAP and \\
    \normalsize Department of Computer Science \\
    \normalsize University of Warwick \\
    \normalsize A.M.Adamaszek@warwick.ac.uk
        \and
    \textbf{Artur Czumaj} \\
    \normalsize DIMAP and \\
    \normalsize Department of Computer Science \\
    \normalsize University of Warwick \\
    \normalsize A.Czumaj@warwick.ac.uk
        \and
    \textbf{Andrzej Lingas} \\
    \normalsize Department of Computer Science \\
    \normalsize Lund University \\
    \normalsize Andrzej.Lingas@cs.lth.se
}

\date{\today}
\date{}

\usepackage{amsmath,amssymb,amsfonts}
\usepackage{xspace}
\usepackage{lettrine}
\usepackage{graphicx}
\usepackage{subfigure}
\usepackage{latexsym}
\usepackage{graphicx}
\usepackage{fancybox}
\usepackage{fullpage}
\usepackage{color}
\usepackage{paralist}
\newcount\shortyear\newcount\shorthour\newcount\shortminute
\shorthour=\time\divide\shorthour by 60\shortyear=\shorthour
\multiply\shortyear by 60\shortminute=\time\advance\shortminute by
-\shortyear \shortyear=\year\advance\shortyear by -1900

\def\zeit{\number\shorthour:\ifnum\shortminute<10 0\number\shortminute
\else\number\shortminute\fi}

\makeatletter

\newenvironment{proof}{\noindent {\bf Proof}.\ }{\qed \par\vskip 4mm\par}

\newcommand{\sq}{\hbox{\rlap{$\sqcap$}$\sqcup$}}
\newcommand{\qed}{\hspace*{\fill}\sq}

\newenvironment{proof-sketch}{\noindent {\bf Sketch of the proof}.\ }{\qed \par\vskip 4mm\par}

\newtheorem{theorem}{Theorem}
\newtheorem{corollary}[theorem]{Corollary}
\newtheorem{lemma}[theorem]{Lemma}

\newtheorem{fact}[theorem]{Fact}{\bfseries}{\itshape}

\newtheorem{cclaim}[theorem]{Claim}

\newcommand{\COMMENTED}[1]{{}}
\newcommand{\junk}[1]{{}}

\renewcommand{\O}{\ensuremath{\mathcal{O}}}

\def\epsilon{\varepsilon}

\newcommand{\opt}{\text{\sc opt}}

\begin{document}

%===============================================================

%\begin{titlepage}

\maketitle
\begin{abstract}
Let $P$ be a set of $n$ points in the Euclidean plane and let $O$ be
the origin point in the plane. In the \emph{$k$-tour cover problem}
(called frequently the \emph{capacitated vehicle routing problem}),
the goal is to minimize the total length of tours that cover all
points in $P$, such that each tour starts and ends in $O$ and covers
at most $k$ points from $P$.

The $k$-tour cover problem is known to be $\mathcal{NP}$-hard. It is
also known to admit constant factor approximation algorithms for all
values of $k$ and even a polynomial-time approximation scheme (PTAS)
for small values of $k$, i.e., $k=\O(\log n / \log\log n)$.

We significantly enlarge the set of values of $k$ for which a PTAS
is provable. We present a new PTAS for all values of $k \le
2^{\log^{\delta}n}$, where $\delta = \delta(\epsilon)$. The main
technical result proved in the paper is a novel reduction of the
$k$-tour cover problem with a set of $n$ points to a small set of
instances of the problem, each with $\O((k/\epsilon)^{\O(1)})$
points.
\end{abstract}

%\end{titlepage}

%===============================================================

\section{Introduction}

The \emph{$k$-tour cover problem} ($k$-TC), is a very natural and
well known generalization of the traveling salesperson problem (TSP)
to include several tours \cite{AKTT96,AKTT97,HK85,TV02}. Namely, we
are given a set $P$ of points (sites), a distinguished point $O$
outside $P$, called the origin as well as a distance function
defined on $P \cup \{ O\}$. A tour is a cycle whose vertices are in
$P \cup \{O\}$. The length of a tour is the sum of distances between
the adjacent points on the tour. The objective is to find a set of
tours, each including the origin and at most $k$ points in $P$,
which covers all points in $P$ and achieves the minimum total
length.

In Operations Research, the $k$-TC problem is well known as the
\emph{capacitated vehicle routing problem} \cite{TV02}. The name
comes from its standard application when the points in $P$ represent
customer locations, and the origin $O$ stands for a depot. Then, a
fleet of vehicles located at the depot must serve all the customers,
so that each vehicle can serve at most $k$ customers. The objective
is to minimize the total distance traveled by the fleet. The $k$-TC
problem (capacitated vehicle routing problem) is one of the central
special cases of a more general vehicle routing problem, introduced
by Dantzig and Ramser \cite{DR59} fifty years ago, and studied very
extensively in the literature ever since (cf. \cite{L92,TV02}).

The $k$-TC problem contains the TSP problem as a special case and it
is known to be $\mathcal{NP}$-hard for all $k \ge 3$. For this
reason, the research on $k$-TC has focused on heuristic algorithms
and approximation algorithms. The most extensively studied variants
of $k$-TC are the metric one, when the distance function is
symmetric and satisfies the triangle inequality, and in particular
the \emph{two-dimensional Euclidean} one, when the points are placed
in the plane and the distance is Euclidean.

The general metric case of $k$-TC for $k \ge 3$ has been shown to be
APX-complete \cite{AKTT96}, i.e., complete for the class of
optimization problems admitting constant factor approximations.
However, the approximability status of the two-dimensional
\emph{Euclidean $k$-TC} problem, in particular, the problem of the
existence of a PTAS, has not been completely settled yet. One of the
first studies of two-dimensional Euclidean $k$-TC has been due to
Haimovich and Rinnooy Kan \cite{HK85}, who presented several
heuristics for the metric and Euclidean $k$-TC, including a PTAS for
the two-dimensional Euclidean $k$-TC with $k < c \log\log n$, for
some constant $c$ \cite[Section~6]{HK85}.
%
%for the Euclidean $k$-TC a factor of $2$ when the more recent PTAS
%for TSP is plugged in \cite{A98,M99}. They also presented the first
%polynomial-time approximation scheme (PTAS) for the two-dimensional Euclidean
%$k$-TC for $k = \O(1)$.
%
Asano et al. \cite{AKTT97} substantially subsumed this result by
designing a PTAS for $k = \O(\log n /\log\log n)$. They also
observed that Arora's \cite{A98,A03} or Mitchell's \cite{M99} PTAS
for the two-dimensional Euclidean TSP implies a PTAS for the
corresponding $k$-TC where $k = \Omega(n)$. There has not been any
significant progress since the paper by Asano et al. \cite{AKTT97}
until very recently, when Das and Mathieu \cite{DM08} showed a
\emph{quasi-polynomial} time approximation scheme (QPTAS) for the
two-dimensional Euclidean $k$-TC for every $k$. Their algorithm
combines the approach developed by Arora \cite{A98} for Euclidean
TSP with some new ideas to deal with $k$-TC (in particular, how to
handle a large number of possible values of the lengths of the
subtours arising in the subproblems of the original $k$-TC), and
gives a $(1+\epsilon)$-approximation for the two-dimensional
Euclidean $k$-TC in time $n^{\log ^{\O(1/\epsilon)} n}$ (this bound
holds for any value of $k$).

In this paper we focus on the two-dimensional Euclidean variant of
$k$-TC. (To simplify the notation, we shall further refer to this
variant as to $k$-TC).

Our \textbf{main result} is a new PTAS for $k$-TC for all values of
$k \le 2^{\log^{\delta}n}$, where $\delta = \delta(\epsilon)$. This
significantly enlarges the set of values of $k$ for which a PTAS is
known. Our PTAS relies on a novel reduction of an instance of $k$-TC
with a set of $n$ points to an instance or a small number of
independent instances of the problem with a small number of points.
Our first reduction takes any instance of $k$-TC on $n$ points and
reduces it to an instance of the problem with
$\O((k/\epsilon)^{\O(1)} \log^2 (n/\epsilon))$ points. Then we
present a refinement, where the instance of $k$-TC is reduced to a
small set of instances of $k$-TC, each with
$\O((k/\epsilon)^{\O(1)})$ points. These results, when combined with
the recent QPTAS due to Das and Mathieu \cite{DM08}, give the
aforementioned PTAS for $k$-TC for all values $k \le
2^{\log^{\delta}n}$, where $\delta = \delta(\epsilon)$.

Our paper is structured as follows. In the next section, we
introduce useful notation and facts regarding $k$-TC. In Section
\ref{sec:PTAS}, we show the first reduction yielding our PTAS. In
Section \ref{sec:refinement}, we present the refined reduction. We
conclude with final remarks.

For simplicity of the presentation, we will present
$(1+\O(\epsilon))$-approxi\-mation algorithms; reduction to $(1 +
\epsilon)$-approximation is straightforward.

%===============================================================

\section{Preliminaries}
\label{section:preliminaries}

We assume a fixed origin in the plane and denote it by $O$. For a
tour $\mathcal{T}$, its (Euclidean) length is denoted by
$|\mathcal{T}|$. For a set $U$ of tours, we set $|U|$ to
$\sum_{\mathcal{T} \in U} |\mathcal{T}|$.

For a set $P$ of points in the plane, we denote by $TSP(P)$ the
minimum length of a TSP-tour through $P$ and by $\opt(P)$ the
minimum length of a solution to $k$-TC (i.e., the minimum length of
a set of tours, each through the origin and containing at most $k$
points of $P$, which covers all points in $P$). When $P$ is clear
from the context, we shall simply use the notation $\opt$.

For a point $p \in P$, we denote by $r(p)$ the distance of $p$ from
the origin~$O$.

The following simple lower bound plays a very important role in the
previous approaches to $k$-TC, see \cite[Proposition~2]{AKTT97} and
\cite[Lemma~1]{HK85}.

\begin{fact}
\label{fact:weak-bound}
    $\opt(P) \ge \frac{2}{k} \sum_{p \in P} r(p)$.
\end{fact}

Following \cite{AKTT97}, we shall term $\frac 2k\sum_{p\in P}r(p)$
as the \emph{radial cost} of $P$, and denote by $rad(P)$. Among
other things, Haimovich and Kan considered the so called
\emph{iterated tour partitioning heuristic} for $k$-TC in
\cite{HK85}. The heuristic starts from constructing a TSP-tour $T$
through $P$. Then, it considers all $k$-tour covers resulting from
partitioning $T$ into paths visiting exactly $k$ points (assuming
that $n$ is divisible by $k$), and connecting the endpoints of the
paths with $O$. The heuristic outputs the shortest among these
solutions.

\begin{fact}{\rm\bf \cite{AKTT97}}
\label{fact:it}
    If the iterated tour partitioning heuristic uses a TSP tour
$U$, then it returns a $k$-tour cover of total length not exceeding
$(1 - \frac{1}{k}) \cdot |U| + rad(P)$.
\end{fact}

Note that given a TSP tour, the iterated tour partitioning heuristic
can be implemented in time $\O(k\frac{n}{k} + n)$ by repeatedly
updating the previous partition and $k$-tour cover to the next one
in time $\O(\frac{n}{k})$. Using the minimum spanning tree heuristic
for TSP we can find a $2$-approximation of the TSP in time $\O(n
\log n)$. Hence, we obtain the following.

\begin{corollary}
\label{cor:mst}
    If the iterated tour partitioning heuristic uses the minimum
spanning tree heuristic for TSP then it returns a $(3-\frac
2k)$-approximation of an optimal $k$-tour cover of an $n$-point set
and it can be implemented in time $\O(n \log n)$.
\end{corollary}

%===============================================================

%\section{Polynomial time approximation scheme for moderate values of $k$}
\section{PTAS for moderate values of $k$}
\label{sec:PTAS}

In this section we present a reduction that takes as an input any
instance of the $k$-tour problem on a set of $n$ points in the
Euclidean plane and reduces it to an instance of the problem with
$\O((k/\epsilon)^{\O(1)} \log^2 (n/\epsilon))$ points. Then, we
apply this reduction to obtain a PTAS for the $k$-tour problem for
all $k \le 2^{\log^{\delta}n}$, where $\delta$ is some positive
constant, $\delta = \delta(\epsilon)$.

Our construction uses a series of transformations that eliminate
most of the input points and reduce the input problem instance to a
significantly smaller one.

%===============================================================

\subsection{Removing close points}

Let $L$ be the maximum distance from a point in $P$ to the origin
$O$, that is, $L = \max\{p \in P : r(p)\}$. Since $\opt \ge 2 L$, we
can ignore any point that is at a distance at most $L\epsilon/n$
from the origin: covering all such points with $1$-tours will give
us additional cost not greater than $n \cdot 2 \frac{L\epsilon}{n}
\le \epsilon \cdot \opt$. Therefore, from now on, we will consider
only the points $p$ with $r(p) \ge L\epsilon/n$.

%===============================================================

\subsection{Circles, rays, and locations}

Let us create \emph{circles} around the origin, the $i$-th circle
with a radius
\begin{eqnarray*}
    c_i & = &
    \frac{L\epsilon}{n} \cdot \left(1 + \frac{\epsilon}{k}\right)^i
            \enspace,
        \qquad
        \textrm{ for } 0 \le i \le
        \left\lceil \log_{(1+\epsilon/k)}\frac{n}{\epsilon} \right\rceil
            \enspace.
\end{eqnarray*}

Let us draw \emph{rays} from the origin with the angle between any
pair of neighboring rays equal to $2 \pi / s$ (that is, partition
the space into $s$ sectors) with $s = \lceil \frac{2 \pi k}
{\epsilon} \rceil$.

%===============================================================
\begin{figure}[t]
\centerline{\includegraphics[width=.8\textwidth]{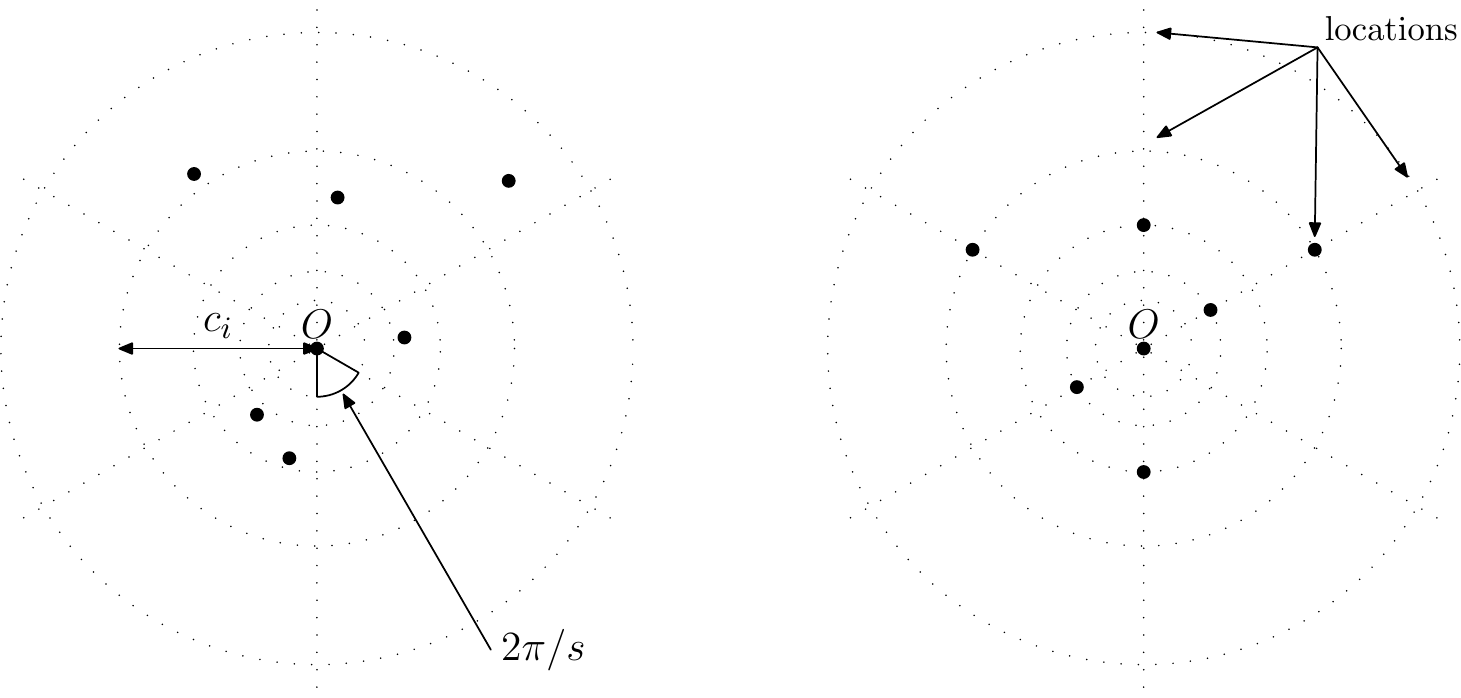}}
\caption{The structure of circles, rays, and locations. The point
labeled $O$ is the origin. Other fat dots represent the points from
$P$. In the right picture each point has been moved to its nearest
location.}
\label{fig:circles-rays-locations}
\end{figure}
%===============================================================

%
Define a \emph{location} to be any point on the plane that is the
intersection of a circle and a ray. Since
\begin{displaymath}
    \log_{(1+\epsilon/k)} \frac{n}{\epsilon}
        =
    \frac{\log \frac{n}{\epsilon}}{\log (1+\epsilon/k)}
        =
    \Theta \left( \frac{k}{\epsilon} \cdot \log \frac{n}{\epsilon} \right)
        \enspace,
\end{displaymath}
there are $\Theta \left(\frac{k}{\epsilon} \log (n/\epsilon)
\right)$ circles and $\Theta \left(\frac{k}{\epsilon} \right)$ rays.
Therefore we obtain:

\begin{cclaim}
\label{claim:num_locations}
    The total number of locations $T$ satisfies $T = \Theta(k^2
\epsilon^{-2} \log(n/\epsilon))$.
\end{cclaim}

Now, we transform the input set $P$ and move each point from $P$ to
its nearest location.

\begin{cclaim}
    The operation of moving each point to its nearest location can
change the cost of a $k$-tour by at most $\epsilon \cdot \opt$.
\end{cclaim}

\begin{proof}
Let $p$ be a point in $P$. Suppose that $p$ lies between the circles
with radius $c_i$ and $c_{i+1}$ (the distance between $p$ and the
origin is in the interval $[L\epsilon/n, L]$, so we know such
circles exist). The distance between these circles equals $c_{i+1} -
c_i = \frac{\epsilon}{k} \cdot c_i$. The distance between two
consecutive locations at the $i$-th circle is less than $2 \pi c_i /
s \le \frac{\epsilon}{k} \cdot c_i$. Therefore the distance between
$p$ and its nearest location is at most $\sqrt{2} \cdot (\frac{1}{2}
\cdot \frac{\epsilon}{k} c_i) < $ $\frac{\epsilon}{k} \cdot c_i \le
\frac{\epsilon}{k} \cdot r(p)$.

If we move a point $p \in P$ by a distance at most
$\frac{\epsilon}{k} \cdot r(p)$, the cost of a tour can change by at
most $2 \frac{\epsilon}{k} \cdot r(p)$. If we add up the changes of
the cost generated by moving all points in $P$, then this total
change is upper bounded by $\sum_{p \in P} 2 \frac{\epsilon}{k}
\cdot r(p)$. Next, we use Fact \ref{fact:weak-bound} to conclude
that the total cost of moving all the points is at most $\epsilon
\cdot \opt$.
%\qed
\end{proof}

From a $k$-tour $U'$ for a modified instance of the problem (where
all points have been moved to their nearest locations) we can easily
get a $k$-tour $U$ for the original version of the problem such that
$|U| \le |U'| + \epsilon \cdot \opt$. So a PTAS for the modified
version yields a PTAS for the original version. In the rest of this
paper we will consider the modified version of the problem.

%===============================================================

\subsection{Trivial and nontrivial tours}

We say that a tour \emph{visits} a location if it contains at least
one point from that location. (If an edge of a tour passes trough a
location, but the tour does not contain any point from that
location, then the tour does not visit that location.)

We call a tour \emph{trivial} if it visits only a single location in
$P$; a tour is \emph{nontrivial} otherwise.

\begin{theorem}
\label{th-T-nontrivial-tours}
    There is an optimal solution in which there are at most $T$
nontrivial tours.
\end{theorem}

\begin{proof}
We say that a set of tours $t_1, t_2, \dots , t_m$ ($m \ge 2$) forms
a \emph{cycle} if there is a set of locations $\ell_1, \ell_2,
\cdots , \ell_m, \ell_{m+1} = \ell_1$ such that each tour $t_i$
visits locations $\ell_i$ and $\ell_{i+1}$. Note that the origin is
not considered as a location.

To prove our theorem we will need the following:

\begin{lemma}
\label{lemma:opt-has-no-cycles}
    There is an optimal solution in which there are no cycles.
\end{lemma}

\begin{proof}
Let $U$ be such an optimal solution which minimizes the sum over all
its nontrivial tours of the number of locations visited by that
tour.

Let us suppose that $U$ has a cycle, and let $t_1, t_2, \dots , t_m$
be a minimal cycle ($m$ is minimal). Let  $\ell_1, \ell_2, \dots ,
\ell_m$ be the locations in which the consecutive tours meet. From
the minimality of the cycle we know that both tours and locations
are pairwise distinct.

Let $v(t,\ell)$ denote the number of points from a location $\ell$
visited by a tour $t$. Let $\min=\min_{i \in \{1, \dots ,m\}}
\{v(t_i,\ell_i)\}$. Now we are ready to swap points between the
tours: the $i$-th tour, instead of visiting $v(t_i,\ell_i)$ points
in the location $\ell_i$ and $v(t_i,\ell_{i+1})$ points in the
location, $\ell_{i+1}$ will now visit $(v(t_i,\ell_i) - \min)$
points in $\ell_i$ and $(v(t_i,\ell_{i+1}) + \min)$ points in
$\ell_{i+1}$. Here $\ell_{m+1}$ denotes $\ell_1$.

Observe that the modification does not change the number of points
visited by each tour. It also does not increase the length of any
tour. Therefore, we obtain another optimal solution, in which the
sum over all nontrivial tours of the number of locations visited by
that tour is smaller than in $U$ (we managed to remove one location
from each tour $t_i$ for which $v(t_i,\ell_i) = \min$). This is a
contradiction with the minimality of that sum in $U$.

Therefore the optimal solution $U$ has no cycles.
%\qed
\end{proof}

Consider an optimal solution without cycles. Note that the lack of
$2$-cycles means that no two tours visit the same pair of locations.
To each nontrivial tour we can assign a pair of distinct locations
visited by this tour. The chosen pairs are in one-to-one
correspondence with the nontrivial tours and they induce an acyclic
undirected graph on the locations.

Hence, we can have at most $T-1$ nontrivial tours in an acyclic
solution, so using Lemma \ref{lemma:opt-has-no-cycles} we have
proved the theorem.
%\qed
\end{proof}

%===============================================================

\subsection{Reduction to an instance of $k$-TC with $(k \log n/\epsilon)^{\O(1)}$ points}

Observe that Theorem \ref{th-T-nontrivial-tours} implies that there
is an optimal solution in which at most $Tk$ points are covered by
nontrivial tours. Therefore it is enough to consider only solutions
which fulfill that property.

If the number of points in a location $\ell$ is greater than $Tk$,
some of the points will have to be covered by trivial tours. We may
assume, without loss of generality, that among all trivial tours
visiting a given location there is at most one that visits less than
$k$ points. Moreover, if at least one point from some location is
visited by a nontrivial tour, we can assume that all trivial tours
visiting that location contain exactly $k$ elements. Therefore, for
each location $\ell$ containing $c_\ell$ points, we only have to
consider at most $\min\{c_\ell, c_\ell - k \cdot \lceil\frac{c_\ell
- Tk}{k}\rceil\} \le Tk$ points for nontrivial tours. After finding
a $(1+\epsilon)$-approximation for such reduced case, we will add
trivial tours covering all remaining points. That will give us
$(1+\epsilon)-$approximation for the original problem.

\begin{corollary}
\label{corollary:1st-reduction}
    One can reduce the $k$-TC problem on $n$ points to one on at
most $T^2k$ points.
\end{corollary}

%===============================================================

%\subsection{Polynomial-time approximation scheme for $k$-TC with $k \le 2^{\log^{\delta} n}$}
\subsection{PTAS for $k$-TC with $k \le 2^{\log^{\delta} n}$}

We use Corollary \ref{corollary:1st-reduction} to reduce any
instance of $k$-TC with the input set of $n$ points $P$ to an
instance of $k$-TC with $N = T^2k = \Theta(k^5 \epsilon^{-4}
\log^2(n/\epsilon))$ input points. For such input instance, we apply
the quasi-polynomial time approximation scheme for $k$-TC due to Das
and Mathieu \cite{DM08}. The obtained algorithm returns a
$(1+\epsilon)$-approximation in time $N^{\log^{\O(1/\epsilon)}N}$.
This gives polynomial time for all $k \le 2^{\log^{\delta} n}$ for
some constant $\delta = \delta(\epsilon) > 0$. Hence, we have the
following main theorem.

\begin{theorem}
\label{main:theorem}
    There is a PTAS for the $k$-TC problem provided that $k \le
2^{\log^{\delta} n}$ for some positive constant $\delta =
\delta(\epsilon)$.
\end{theorem}

%===============================================================

\section{Refinement: reduction to $(k/\epsilon)^{\O(1)}$ points}
\label{sec:refinement}

In the preceding section, we have demonstrated that the problem of
close approximation of the $k$-TC problem on the input set of $n$
points in the plane reduces to that for a multi-point-set of size
polynomial in $k/\epsilon$ and polylogarithmic in $n$ in the
relevant locations. In this section, we shall eliminate the
polylogarithmic dependency of $n$ in the reduction. This will have
only a relatively small effect on the asymptotics for the size of
the largest $k$ in terms of $n$ for which we can attain a PTAS and
we will obtain a PTAS for all $k \le 2^{\log^{\delta'} n}$, where
comparing to the bound in Theorem \ref{main:theorem}, we will have
$\delta' > \delta$. However, for small values of $k$ this will lead
to a faster PTAS. Hopefully, because it removes completely the
dependency on $n$ from the size of the reduced instance, it also
might be a step towards a PTAS for arbitrary values of $k$.

The idea of our refinement resembles Baker's method \cite{B94} of
closely approximating several hard problems on planar graphs. It
relies on the following separation lemma.

%Further, we shall denote by $r_i$ the $i$-th ring for $i=1, \dots $.

\begin{lemma} \label{lem:sep}
    Let $P$ be a set of points situated in the locations and
let $\epsilon > 0$. There is a clustering of the circles into rings
of $\lceil \log_{1+\frac {\epsilon}k}(6/\epsilon)\rceil$ consecutive
circles and there are positive integers $a = \O(\epsilon^{-1})$ and
$b \in \{ 1, \dots , a\}$ such that if we mark each $(b+ja)$-th ring
then any $k$-tour cover $U$ of $P$ can be transformed to a $k$-tour
cover $U'$ of the points in the unmarked rings such that
\begin{enumerate}
\item no tour in $U'$ visits two points in $P$ separated by a
    marked ring, and
\item $|U'| \le (1+\frac{\epsilon}{2})|U|$.
\end{enumerate}
Furthermore, the points in the marked rings can be covered with
$k$-tours of total length at most $\frac {\epsilon}2 |U|$ produced
by the iterated tour partitioning heuristic from \cite{HK85} (cf.
Section \ref{section:preliminaries}).
%, using as a subroutine the minimum spanning tree heuristic for TSP.
\end{lemma}

\begin{proof}
Let $t$ denote a tour obtained by removing its edges incident to
$O$. Suppose that $t$ crosses one of the marked rings. Let $i$ be
the number of the most inner circle of the ring. Denote the circle
by $C_i$. It follows by straightforward calculation and the
definition of the circles that each minimal fragment of $t$ crossing
the aforementioned ring is at least $\frac{2}{\epsilon}$ times
longer than the doubled radius of $C_i$. We can appropriately split
the tour $t$ along $C_i$ into smaller ones by connecting pairs of
crossing points on $C_i$ with $O$ or just with themselves, see
Figure \ref{fig:Haimovich-RinnoyKann}.

%===============================================================
\begin{figure}[t]
\centerline{\includegraphics[width=.7\textwidth]{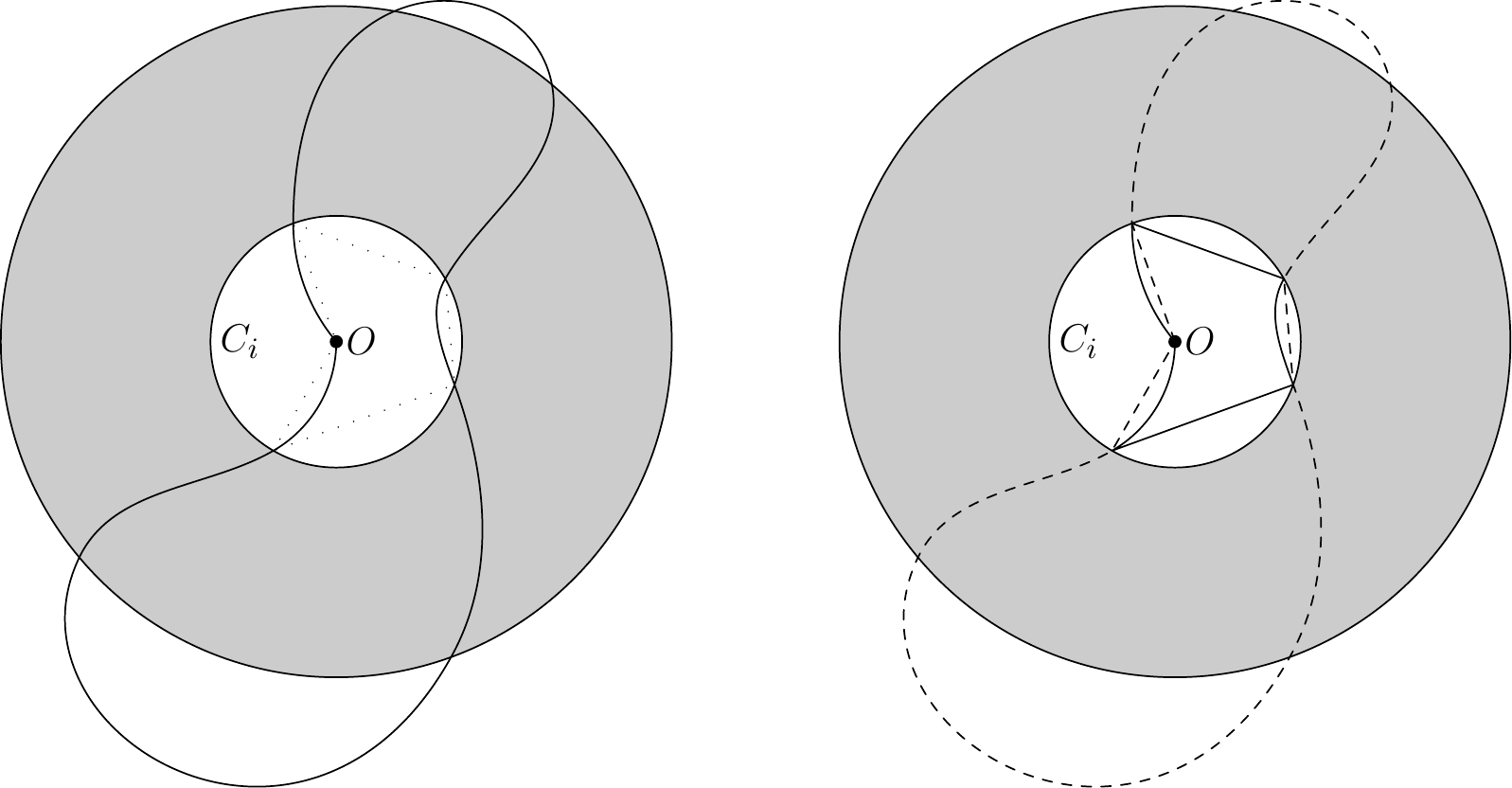}}
\caption{Splitting $t$ into smaller tours. The grey area is the
marked ring. In the left picture dotted lines represent the lines
which will be added to our solution. The right picture shows two
separate tours obtained from the original tour (one is marked with a
dashed line, and the other with a solid one), before the
short-cutting.}
\label{fig:Haimovich-RinnoyKann}
\end{figure}
%===============================================================

The total length of the smaller tours is longer than $|t|$ by at
most $\frac{\epsilon}{2}$ of the total length of the aforementioned
fragments of $t$.

We may assume, without loss of generality, that the aforementioned
marked ring is the outermost among those crossed by $t$. We can
iterate the elimination of the crossings of the smaller resulting
tours but for their edges incident to $O$ with more inner marked
rings. Note that then other disjoint fragments of $t$ will be
charged with the increase of the length of the union of the
resulting smaller tours. Finally, by applying short-cutting, we can
drop the points in the marked rings from the resulting tours.

We conclude that we can transform $U$ into a $k$-tour cover $U'$ of
the points in $P$ in the unmarked rings such that no tour in $U'$
crosses any marked ring (but for its edges incident to $O$) and
$|U'|\le (1+\frac {\epsilon}2)|U|$.

It remains to show that we can set $a$ and $b \in \{ 1, \dots , a\}$
such that one can easily cover the points in $P$ contained in the
marked rings with $k$-tours of total length not exceeding
$\frac{\epsilon |U|}{2}$.

%For a set $P$ of points, let $R(P)=\frac 2 k\sum_{p\in P} dist(p,O)$,
%let $TSP(P)$ be the minimum length of a tsp tour on $P$, and let $k-TC(P)$
%be the length of minimum $k$-tour cover of $P$ (we assume that $O$ is fixed).
%By \cite{}, $k_TC(P)\le R(P) + TSP(P)$.

Let $R_j$ denote the set of points from $P$ lying in the $j$-th
ring. Set $a$ to $\lceil \frac{24}{\epsilon}\rceil $. For each $b
\in \{ 1, \dots ,a \}$, let $P_b$ be the set of points in $P$ in the
marked rings, $P_b = \sum_{j \equiv b \bmod a}R_j$. We shall show
that there is some $b \in \{ 1, \dots , a\}$ such that by applying
the $k$-TC heuristic given in Corollary \ref{cor:mst} for $P_b$, we
can cover $P_b$ with $k$-tours of length at most $\frac{\epsilon
|U|}{2}$. For this purpose, we shall observe that $\sum_j TSP(R_j)
\le 3 \cdot TSP(P)$.

Suppose for the sake of this observation that the tour $t$
considered in the first part of the proof is an $n$-tour, i.e., an
optimal TSP tour of $P \cup \{ O\}$. Apply almost the same
transformation to the tour $t$ as before with the exception that
instead of connecting the outer cut part by two rays to $O$, we
connect the cutting points directly. By the triangle inequality, the
total length of the so modified TSP tour $t$ is at most $(1 +
\frac{\epsilon}{2}) \cdot TSP(P)$. The modified TSP tour $t$ can be
easily reduced to the non-necessarily optimal TSP tours of the
unmarked regions by short-cutting. Assuming first for a moment that
the unmarked rings are the even ones, and then conversely, that the
unmarked rings are the odd ones, and that $\epsilon < \frac12$, we
conclude that $\sum_j TSP(R_j) \le 3 \cdot TSP(P)$.

Using Fact \ref{fact:it} we get that
\begin{eqnarray*}
    \sum_{b \in \{1, \dots ,a\}} \opt(P_b)
        & \le &
    \sum_{b \in \{1, \dots ,a\}} \sum_{j \equiv b(mod\ a)} \opt(R_j)
        \\
        & = &
    \sum_j \opt(R_j)
        \\
        & \le &
    \sum_j (rad(R_j) + TSP(R_j))
        \\
        & \le &
    rad(P) + 3 \cdot TSP(P)
        \\
        & \le &
    4 |U|
        \enspace.
\end{eqnarray*}

There must be some $b \in \{ 1, \dots , a\}$ such that $\opt(P_b)
\le \frac{4}{a} |U| \le \frac{\epsilon |U|}{6}$. Thus, if we apply
the $3$-approximation algorithm for the $k$-tour of $P_b$, which is
a composition of the iterated tour partitioning heuristic with the
minimum spanning tree heuristic for TSP, we obtain a $k$-tour cover
of $P_b$ of length not exceeding $\frac{\epsilon |U|}{2}$.
%\qed
\end{proof}

\begin{theorem}\label{theo: ref}
    The $k$-TC problem on a set $P$ of $n$ points on the plane can
be reduced to a collection of $\O(\epsilon^{-1} \log (n/\epsilon) /
\log (1/\epsilon))$ disjoint $k$-tour cover problems, each on
$\O(k^5 \epsilon^{-6} \log^2(1/\epsilon))$-point set and each having
the maximum distance to the origin at most $(1/\epsilon)^{\O(1 /
\epsilon)}$ larger than the minimum one, such that
$(1+\epsilon)$-approximate solutions to each of the latter problems
yield a $(1+\O(\epsilon ))$-approximation to the original $k$-tour
cover problem. The reduction can be done in time $\O(n \log n)$ for
a fixed $\epsilon$.
\end{theorem}

\begin{proof}
Move the points to the locations and compute the sets $R_j$ of input
points lying in the rings for a fixed $\epsilon$. This all can be
easily done in time $\O(n\log n)$ by using standard data structures
for point location \cite{PS85}.

Next, compute the value $a$ (the distance between marked rings) and
for each $b \in \{ 1, \dots , a\}$, compute a $3$-approximate
$k$-tour cover of the set $P_b$ of points contained in the marked
rings. All the $a$ computations take $\O(an \log n) = \O(n \log n)$
time by Corollary \ref{cor:mst}.

Fix $b$ to that minimizing the length of the aforementioned tour. It
follows from Lemma \ref{lem:sep} that the produced cover of $P_b$
has length at most $\frac{\epsilon}{2} \opt$. Now we will have to
compute approximate solutions for each maximal sequence of
consecutive not marked rings. Let us denote the number of such
sequences by $q$. We can easily compute that $q=\O(\epsilon^{-1}
\log \frac{n}{\epsilon} / \log \frac{1}{\epsilon})$. For $i = 1,
\dots , q$, let $I_i$ denote the set of points contained in such
$i$-th sequence. Note that these point sets can be also easily
computed in time $\O(n \log n)$.

It follows from Lemma \ref{lem:sep} that if we compute separately
$(1+\epsilon)$-approximation of the optimal cover with $k$-tours for
each set $I_i$, then the union of these coverings will have length
at most $(1+\O(\epsilon)) \opt$.

Note that for a given $i$, the number of locations in $I_i$ is $\O(a
\cdot \frac{k}{\epsilon} \cdot \log_{(1+\frac
{\epsilon}k)}\frac{1}{\epsilon}) = \O(k^2\epsilon^{-3}\log
\frac{1}{\epsilon})$. Hence, by the discussion in Section
\ref{sec:PTAS}, we can account to the intended
$(1+\epsilon)$-approximation of $\opt(I_i)$ the trivial tours
decreasing the point-multiplicity in each location to $\O(k^3
\epsilon^{-3} \log \frac{1}{\epsilon})$. Thus, for each $I_i$ we can
reduce the problem to one with $\O(k^5\epsilon^{-6}(\log
\frac{1}{\epsilon})^2)$ points.

Each $I_i$ consists of $\O(\epsilon^{-1})$ consecutive rings and for
a point in a ring the maximum distance to the origin is at most
$\O(\epsilon^{-1})$ times larger than the minimum one. Hence, for a
point in $I_i$ the maximum distance to the origin is at most
$(1/\epsilon)^{\O(1/\epsilon)}$ times larger than the minimum one.

The appropriate $q$ sets of points can be computed in time $\O(n
\log n)$ and they specify the problems to which we approximately
reduce the original $k$-tour cover problem.
%\qed
\end{proof}

%===============================================================

%\section{Conclusions}
\section{Final remarks}

In this paper, we have considered the problem of approximating
two-dimensional Euclidean $k$-TC. Prior to our work, a PTAS has been
known only for the values of $k \le \O(\log n / \log\log n)$ and for
$k = \Omega(n)$ \cite{AKTT97}, and in this paper we significantly
enlarge the set of values of $k$ to $k \le 2^{\log^{\delta}n}$ for
some positive constant $\delta = \delta(\epsilon)$. The main
technical contribution is a reduction of the $k$-TC problem on $n$
points to either that on $(k \log n/\epsilon)^{\O(1)}$ points, or to
a small number of independent instances of the $k$-TC problem on
$(k/\epsilon)^{\O(1)}$ points. When combined with a QPTAS for $k$-TC
due to Das and Mathieu \cite{DM08}, this gives a PTAS for $k \le
2^{\log^{\delta}n}$ for some positive constant $\delta =
\delta(\epsilon)$.

The central open question left is whether there is a PTAS for the
$k$-TC problem for all values of $k$. While we have enlarged the set
of values of $k$ for which a PTAS exists, we still do not know how
to reach polynomial values for $k$, even $k = n^{0.001}$. In
particular, a PTAS $k$-TC for $k = \Theta(\sqrt{n})$ is elusive. For
arbitrary values of $k$, the best currently known result is either a
quasi-polynomial time approximation scheme by Das and Mathieu
\cite{DM08} that runs in time $n^{\log^{\O(1/\epsilon)}n}$, or the
polynomial-time constant-factor approximation algorithm due to
Haimovich and Rinnooy Kan \cite{HK85}. Similarly as in
\cite{AKTT97}, we believe that the case $k = \Theta (\sqrt{n})$ is
the hardcore of the difficulty in obtaining a PTAS for all values of
$k$.

Following \cite{HK85}, let us observe that if we divide the range of
$k$, i.e., the interval $\{1, \dots, n\}$, into a logarithmic number
of intervals of the form $[\epsilon^{-2i}, \epsilon^{-2(i+1)})$,
then for $k$ in at most one of the intervals none of the
inequalities $TSP(P)\le \epsilon \cdot rad(P)$, $rad(P)\le \epsilon
\cdot TSP(P)$ hold. Note that if any of the inequalities holds then
by plugging any PTAS for TSP in the iterated tour partitioning
heuristics, we obtain an $(1+\O(\epsilon))$-approximation of $k$-TC.
Thus, the aforementioned heuristic is in fact a PTAS for a
substantial range of $k$ depending on $P$: for every set of points
$P$ there is $k_0$ such that there is a polynomial-time
$(1+\O(\epsilon))$-approximation algorithm for $k$-TC for every $k
\le \epsilon k_0$ and for every $k > k_0/\epsilon$.
%
%However, even though for every $P$, this implies a PTAS for $k$-TC
%for \emph{almost all} values of $k$,
Despite this observation and despite recent progress in
\cite{AKTT97,DM08}, the problem of designing a PTAS for \emph{all} $k$
remains open: we believe that our paper sheds the light on this
problem and is a step towards a PTAS for arbitrary values of
$k$.

%===============================================================

\newcommand{\CIAC}{Italian Conference on Algorithms and Complexity}
\newcommand{\COCOON}{Annual International Computing Combinatorics Conference (COCOON)}
\newcommand{\COMPGEOM}{Annual ACM Symposium on Computational Geometry (SoCG)}
\newcommand{\ESA}{Annual European Symposium on Algorithms (ESA)}
\newcommand{\FOCS}{IEEE Symposium on Foundations of Computer Science (FOCS)}
\newcommand{\FSTTCS}{Foundations of Software Technology and Theoretical Computer Science (FSTTCS)}
\newcommand{\ICALP}{Annual International Colloquium on Automata, Languages and Programming (ICALP)}
\newcommand{\IPCO}{International Integer Programming and Combinatorial Optimization Conference (IPCO)}
\newcommand{\ISAAC}{International Symposium on Algorithms and Computation (ISAAC)}
\newcommand{\ISTCS}{Israel Symposium on Theory of Computing and Systems}
\newcommand{\JACM}{Journal of the ACM}
\newcommand{\LNCS}{Lecture Notes in Computer Science}
\newcommand{\MOR}{Mathematics of Operations Research}
\newcommand{\SICOMP}{SIAM Journal on Computing}
\newcommand{\SIJDM}{SIAM Journal on Discrete Mathematics}
\newcommand{\SODA}{Annual ACM-SIAM Symposium on Discrete Algorithms (SODA)}
\newcommand{\SPAA}{Annual ACM Symposium on Parallel Algorithms and Architectures (SPAA)}
\newcommand{\STACS}{Annual Symposium on Theoretical Aspects of Computer Science (STACS)}
\newcommand{\STOC}{Annual ACM Symposium on Theory of Computing (STOC)}
\newcommand{\SWAT}{Scandinavian Workshop on Algorithm Theory (SWAT)}
\renewcommand{\SWAT}{SWAT}
\newcommand{\TCS}{Theoretical Computer Science}

\newcommand{\Proc}{Proceedings of the }
%\renewcommand{\Proc}{Proc. }

%===============================================================

\end{document}